\documentclass{elsart}
\usepackage{graphicx}

\begin{document}

\begin{frontmatter}

\title{EPR correlations and EPW distributions revisited}

\author{Lars M. Johansen \thanksref{Email} \thanksref{Permanent}}
\address{Institute of Physics, University of Oslo, P.O.Box 1048
Blindern, N-0316 Oslo, Norway}
\date{\today}

\thanks[Email]{Email: lars.m.johansen@hibu.no}
\thanks[Permanent]{Permanent address: Buskerud College,
P.O.Box 251, N-3601 Kongsberg, Norway}

\begin{abstract}

It is shown that Bell's proof of violation of local realism in phase
space is incorrect. Using Bell's approach, a violation can be derived
also for nonnegative Wigner distributions. The error is found to lie
in the use of an unnormalizable Wigner function.

\end{abstract}


\end{frontmatter}

In 1964, J. S. Bell derived an inequality which must be obeyed by any
local, realistic theory \cite{Bell64}. Simultaneously, he
demonstrated that quantum mechanics violates such an inequality. This
was done using a two-particle entangled spin-state. Clauser \emph{et
al.} reformulated Bell's inequality in a way better adapted to
experimental testing \cite{Clauser69}. Since then, Bell's theorem has
been extended in many ways \cite{Clauser74b,Gisin92}. Theoretically,
violation of local realism has also been demonstrated without the use
of inequalities \cite{Greenberger89,Hardy92}. Local realism has been
tested in various experiments, and in the overwhelming majority of
cases a violation has been found \cite{Aspect81,Chiao95}.

Most of this activity has been concerned with tests where discrete
variables are measured. Most typically, entangled spin-$1 \over 2$
states \cite{Clauser78,Pipkin78} are employed. Bell's paper
\cite{Bell64} was inspired by the work of Einstein, Podolsky and
Rosen (EPR), which dealt with a gedanken experiment where
\emph{continuous} variables like position and momentum would be
observed \cite{Einstein35}. In 1985, Horne and Zeilinger proposed a
Bell-type experiment using entanglement in continuous variables, in
this case linear momenta \cite{Horne85}. This possibility was
elaborated further in Refs. \cite{Zukowski88,Horne89}, and an
experimental test was performed by Rarity and Tapster in 1990
\cite{Rarity90}.

It has been shown that the EPR state can be generated in quantum
optics by a nondegenerate parametric amplifier \cite{Reid88,Reid89}.
In this experiment, two conjugate variables can both be observed with
arbitrary precision if one of them is \emph{inferred} from a strongly
correlated variable. These predictions were later confirmed in an
experiment by Ou {\em et al.} \cite{Ou92a,Ou92b}. It should be noted
that violation of local realism is not demonstrated in such
experiments.

In 1986 Bell presented a paper entitled ``EPR correlations and EPW
distributions" in a conference arranged by the New York Academy of
Sciences \cite{Bell86}. It has also been reprinted in the volume
``Speakable and unspeakable in quantum mechanics" \cite{Bell87a}.
In the title, Bell plays with the acronyms EPR and EPW, the latter
being formed from the initials of Eugene P. Wigner. The Wigner
distribution \cite{Wigner32} plays a prominent role in this paper.
With a few notable exceptions \cite{Leonhardt93b,Leonhardt95b}, the
paper has received little attention in the litterature. It is
different from the other works by Bell on local realism in that it
deals with an experiment where variables with a continuous spectrum
are to be observed. Bell formulates a condition on local realism for
a two-particle system where the particle positions are to be observed
at different times. This is closely related to the original EPR
problem. The state originally used by EPR has a nonnegative Wigner
distribution. Bell argues that this constitutes a local, classical
model, and that consequently this state should not violate local
realism. He then examines a state with negative Wigner distribution,
and proceeds to demonstrate that this state violates local realism.

Recently, Leonhardt and Vaccaro \cite{Leonhardt95b} have demonstrated
that Bell's scheme can be translated into a quantum optical
experiment. Instead of observing particle positions at different
times, they suggested to observe the rotated quadrature variables of
a two-mode radiation field at different settings of the local
oscillator phases. As the input state, they suggested to use a
superposition of vacuum and a two-photon state.

In his 1986 paper \cite{Bell86}, Bell used an unnormalizable
Wigner function in order to describe a state with sharp momentum.
However, the use of vectors with infinite norm may sometimes lead to
problems \cite{Peres93}. Due to the infinite norm, Bell could not
calculate normalized probabilities. Instead, he introduced an
inequality where only relative probabilities enter. With these
assumptions he was able to demonstrate violation of local realism. In
this paper, I shall follow Bell's procedure in Ref. \cite{Bell86}
closely, but I will use an input state which has an explicitly
nonnegative Wigner distribution. Nevertheless, I will derive a
violation of Bell's inequality. This raises the suspicion that there
must be something wrong in Bell's argument in Ref. \cite{Bell86}. And
indeed, by closer inspection, it is found that the use of an
unnormalizable Wigner function is not permissible, due to the fact
that a normalization factor is treated as time independent. This
will be explained in detail later.

To begin with, we consider a coherent state $\mid \alpha \rangle$.
Here $\alpha= (1/\sqrt{2}) (q_0+i p_0)$, and $q_0$ and $p_0$ are
assumed to be real numbers. This state has a Wigner distribution
\cite{Wigner32,Walls94}
\begin{equation}
	W_1(q,p) = {1 \over \pi} \exp \left [ -(q-q_0)^2 - (p-p_0)^2
	\right ].
	\label{eq:seed}
\end{equation}
Also, we shall consider a ``squeezed vacuum" state, which has a
Wigner distribution \cite{Walls94}
\begin{equation}
	W_2(Q,P) = {1 \over \pi} \exp \left [ - (s Q)^2 \right ] \exp
	\left [ - \left ({P \over s} \right )^2 \right ].
	\label{eq:squeezed}
\end{equation}
When $s$ decreases, the first gaussian gets very broad and almost
independent of $Q$, whereas the second gaussian gets very narrow.
In the limit $s \rightarrow 0$ Bell writes the Wigner function as
\begin{equation}
	W_2(Q,P) = K \; \delta(P),
	\label{eq:delta}
\end{equation}
where he denotes $K$ as an ``unimportant constant" \cite{Bell86}.
Note that in this way he has started working with an unnormalizable
Wigner distribution. It is the main purpose of this paper to show
that this trick is not permissible.

\begin{figure}
	\centering
	\includegraphics[width=12cm]{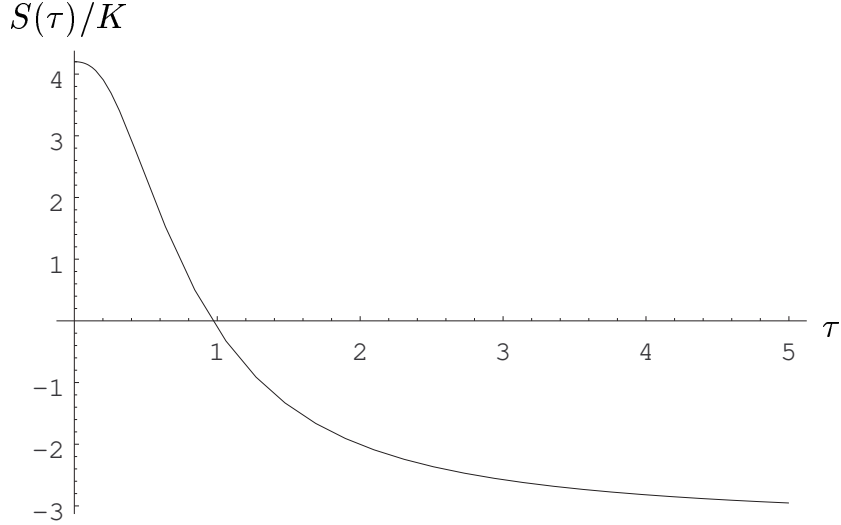}
	\caption{The function $S(\tau)/K$ for the state (\ref{eq:total})
	when $q_0=1$ and $p_0=-1$. It clearly can become negative, a
	result which should not be possible for a nonnegative Wigner
	distribution. This indicates a fault in Bell's argument.}
	\label{fig:neg}
\end{figure}

Assume that the total system is described by the product $W_1 W_2$.
Note that both $W_1$ and $W_2$ are nonnegative. In this sense, a
classical phase space description exists. In analogy with Bell's
treatment (see also Bohr in Ref. \cite{Bohr35}), we impose the linear
transformations
\begin{eqnarray}
	q &=& {1 \over \sqrt{2}} (q_1-q_2), \quad p = {1 \over \sqrt{2}}
	(p_1-p_2), \label{eq:qtrans} \\
	Q &=& {1 \over \sqrt{2}} (q_1+q_2), \quad P = {1 \over \sqrt{2}}
	(p_1+p_2). \label{eq:Ptrans}
\end{eqnarray}
Expressed in terms of output variables, the total Wigner distribution
now reads
\begin{equation}
	W(q_1,q_2,p_1,p_2) = W_1({q_1-q_2 \over \sqrt{2}},{p_1-p_2 \over
	\sqrt{2}})W_2({q_1+q_2 \over \sqrt{2}},{p_1+p_2 \over \sqrt{2}}).
	\label{eq:total}
\end{equation}
Now we assume that the two ``positions" $q_1$ and $q_2$ are measured
at arbitrary times $t_1$ and $t_2$, respectively. The times $t_k$
play the role of local parameters, just like polarizer settings do in
the traditional Bell experiment \cite{Bell64}. For two free
particles, the two-time Wigner distribution is \cite{Wigner32}
\begin{equation}
	W(q_1,q_2,p_1,p_2,t_1,t_2) = W(q_1 - p_1 t_1,q_2 - p_2
	t_2,p_1,p_2).
\end{equation}
Note that this distribution is also nonnegative. We assume that the
two masses are equal, and choose a unit of mass so that $m_k=1$. The
two-time, marginal, joint position probability distribution is
\begin{equation}
	w(q_1,q_2,t_1,t_2) = \int_{-\infty}^{\infty} \d p_1
	\int_{-\infty}^{\infty} \d p_2 \; W(q_1,q_2,p_1,p_2,t_1,t_2).
\end{equation}
Due to the $\delta$ function (\ref{eq:delta}), the first integral
over, say, $p_2$, essentially consists in replacing $p_2$ with
$-p_1$. After performing the final integral over $p_1$, we get
\begin{equation}
	w(q_1,q_2,t_1,t_2) = w(q,\tau),
\end{equation}
where
\begin{equation}
	w(q,\tau) = {K \over \sqrt{\pi}} {1 \over \sqrt{1 + \tau^2}} \exp
	\left [ -{(q - q_0(\tau))^2 \over 1 + \tau^2} \right ]
\end{equation}
and
\begin{equation}
	\tau = {t_1 + t_2 \over 2}, \quad q_0(\tau) = q_0 + p_0 \tau.
\end{equation}
Following Bell (see also Ref. \cite{Leonhardt95b}), we calculate the
unnormalized ``probability" of $q_1$ and $q_2$ having different
signs,
\begin{eqnarray}
	D(t_1,t_2) &=& \int_0^{\infty} \d q_1 \int_{-\infty}^0 \d q_2
	\; w(q_1,q_2,t_1,t_2) \nonumber \\ &+& \int_{-\infty}^0 \d q_1
	\int_0^{\infty} \d q_2 \; w(q_1,q_2,t_1,t_2).
\end{eqnarray}
By change of integration variables using Eqs.
(\ref{eq:qtrans})-(\ref{eq:Ptrans}), we may write
\begin{eqnarray}
	D(t_1,t_2) = F(\tau) &=& \int_{-\infty}^0 \d q \int_{q}^{-q} \d Q
	\; w(q,\tau) + \int_0^{\infty} \d q \int_{-q}^q \d Q \; w(q,\tau)
	\nonumber \\ &=& 2 \int_0^{\infty} \d q \; q \;
	w(q,\tau) - 2 \int_{-\infty}^0 \d q \; q \; w(q,\tau) \nonumber
	\\ &=& 2 \int_0^{\infty} \d q \; q \; \left [ w(q,\tau) +
	w(-q,\tau)
	\right ].
\end{eqnarray}
The result is
\begin{equation}
	F(\tau) = {2 K \sqrt{1+\tau^2} \over \sqrt{\pi}} \exp \left
	[-{q_0(\tau)^2 \over 1+\tau^2} \right ] + 2 q_0(\tau) \;
	\textrm{erf} \left ( {q_0(\tau) \over \sqrt{1 + \tau^2}} \right
	).
\end{equation}
Bell proceeded to demonstrate \cite{Bell86} that for local hidden
variable theories, the function
\begin{equation}
	S(\tau) = 3 F(\tau) - F(3 \tau)
\end{equation}
should be nonnegative,
\begin{equation}
	S(\tau) \ge 0.
	\label{eq:bell}
\end{equation}
By setting, e.g., $q_0=1$ and $p_0=-1$, it is seen that the state
(\ref{eq:total}) violates this inequality, even though it has a
nonnegative Wigner distribution (see Fig. \ref{fig:neg}). How can
this be possible?

According to Bell, $F(\tau)$ expresses the probability of $q_1$
and $q_2$ having different signs, apart from an ``unimportant
constant" $K$ \cite{Bell86}. However, this conclusion is not
straightforward, due to the following argument which applies both to
the state considered by Bell in Ref. \cite{Bell86} and the state
considered here. A probability can never exceed unity, but we see
that $F(\tau)\rightarrow \infty$ as $\tau \rightarrow \infty$. Thus,
in order to have a finite probability as $t \rightarrow \infty$, the
normalization constant $K$ must approach infinity as $t \rightarrow
\infty$. If we now assume that the normalization constant $K$ is time
independent, the probability will vanish for all finite times. This
is clearly wrong, and besides does not lead to any violation of
Bell's inequality (\ref{eq:bell}). We therefore see that the
normalization constant $K$ must depend on time. In the inequality
(\ref{eq:bell}), the probability is compared at two different times,
and one is not allowed to assume that $K$ is time independent. The
conclusion is that the assumption that $K$ is an ``unimportant
constant" is wrong, and that violation of local realism has not been
demonstrated in Ref. \cite{Bell86}.

To summarize, it has been found that Bell's demonstration in Ref.
\cite{Bell86} of violation of local realism in phase space for
continuous variables is untenable. Bell's calculation was repeated,
but with a nonnegative Wigner function. Nevertheless, it was
seemingly possible to derive a violation of local realism. This
should not be possible, since a nonnegative Wigner function can be
regarded as a local hidden variable theory. It was seen that the
error was caused by the use of an unnormalizable state. In this
experiment, the two ``local times" $t_k$ play the role of local
parameters in the same way that the orientation of the spin-filters
are the local parameters in the standard Bell experiment
\cite{Bell64}. It is not permissible to disregard a normalization
constant which depends on the local parameters.

\end{document}